\documentclass[preprint]{aastex}\usepackage{emulateapj5,psfig}
\usepackage{apjfonts}\tighten

\newcommand\etal{{et~al.}} 
\newcommand\mM{\ifmmode(m{-}M)\else$(m{-}M)$\fi}
\newcommand\msun{\ifmmode{\hbox{M$_\odot$}}\else{M$_\odot$}\fi}

\newcommand\hst{{\it HST}}

\newcommand\iz{\ifmmode(i_{775}{-}z_{850})\else$(i_{775}{-}z_{850})$\fi}
\newcommand\izacs{\ifmmode(i_{775}{-}z_{850})\else$(i_{775}{-}z_{850})$\fi}
\newcommand\ubrest{\ifmmode(U{-}B)_z\else$(U{-}B)_z$\fi}
\newcommand\zacs{\ifmmode z_{850}\else$z_{850}$\fi}
\newcommand\iacs{\ifmmode i_{775}\else$i_{775}$\fi}
\newcommand\clname{RDCS\,1252}
\newcommand\zf{\ifmmode z_{\rm f}\else$z_{\rm f}$\fi}
\def\lta{\mathrel{\hbox{\rlap{\hbox{\lower3pt\hbox{$\sim$}}}\raise2pt\hbox{$<$}}}}
\def\gta{\mathrel{\hbox{\rlap{\hbox{\lower3pt\hbox{$\sim$}}}\raise2pt\hbox{$>$}}}}
\newcommand\taul{\ifmmode\tau_{L}\else$\tau_{L}$\fi}

\shortauthors{Blakeslee et al.}
\shorttitle{Color-Magnitude Relation at $z=1.24$}
\slugcomment{To appear in ApJ Letters, 20 October 2003.}
\begin{document}

\title{Advanced Camera for Surveys Photometry of the Cluster RDCS\,1252.9$-$2927:\\
The Color-Magnitude Relation at $z=1.24$ ~~}

\author{John P.~Blakeslee\altaffilmark{1},
Marijn Franx\altaffilmark{2},
Marc Postman\altaffilmark{1,3},
Piero Rosati\altaffilmark{4},
Brad P.~Holden\altaffilmark{5} 
G.~D.~Illingworth\altaffilmark{5},
H.~C.~Ford\altaffilmark{1},
N.~J.~G.\ Cross\altaffilmark{1},
C.~Gronwall\altaffilmark{6}, 
N.~Ben\'{\i}tez\altaffilmark{1},
R.~J.~Bouwens\altaffilmark{1},
T.~J.~Broadhurst\altaffilmark{7},
M.~Clampin\altaffilmark{3},
R.~Demarco\altaffilmark{4},
D.~A.~Golimowski\altaffilmark{1},
G.~F.~Hartig\altaffilmark{3},
L.~Infante\altaffilmark{9}
A.~R.~Martel\altaffilmark{1},
G.~K.~Miley\altaffilmark{2},
F.~Menanteau\altaffilmark{1},
G.~R.~Meurer\altaffilmark{1},
M.~Sirianni\altaffilmark{1}, 
R.~L.~White\altaffilmark{1,3}
}

\altaffiltext{1}{Department of Physics \& Astronomy, Johns Hopkins University, Baltimore, MD 21218; jpb@pha.jhu.edu}
\altaffiltext{2}{Leiden Observatory, P.O. Box 9513, 2300 Leiden, The Netherlands}
\altaffiltext{3}{Space Telescope Science Institute, 3700 San Martin Drive, Baltimore, MD 21218}
\altaffiltext{4}{European Southern Observatory, Karl-Schwarzschild-Str. 2, D-85748 Garching, Germany}
\altaffiltext{5}{Lick Observatory, University of California, Santa Cruz, CA 95064}
\altaffiltext{6}{Department of Astronomy \& Astrophysics, The Pennsylvania State University, University Park, PA 16802.}
\altaffiltext{7}{The Racah Institute of Physics, Hebrew University, Jerusalem 91904, Israel}

\begin{abstract}
We investigate the color-magnitude (CM) relation  of 
galaxies in the distant X-ray selected cluster RDCS\,1252.9--2927 at $z{\,=\,}1.24$
using images obtained with the Advanced Camera for Surveys (ACS)
on the {\it Hubble Space Telescope} in the F775W and F850LP bandpasses.
We select galaxies based on  morphological classifications 
extending about 3.5 mag down the galaxy luminosity function,
augmented by spectroscopic membership information.
At the core of the cluster is an extensive early-type galaxy
population surrounding a central pair of galaxies 
that show signs of dynamical interaction.
The early-type population defines a tight sequence in the CM diagram,
with an intrinsic scatter in observed \iz\ of $0.029\pm0.007$ mag
based on 52 galaxies, or $0.024\pm0.008$ mag for $\sim\,$30 ellipticals.
Simulations using the latest stellar population models
indicate an age scatter for the ellipticals of about 34\%,
with a mean age $\taul\gta2.6$ Gyr (corresponding to $z_L\gta2.7$),
and the last star formation occurring at $z_{\rm end}\gta1.5.$ 
Transforming to rest-frame $(U{-}B)$,
we conclude that the slope and scatter in the CM relation for 
morphologically selected early-type galaxies show little or no evidence
for evolution out to $z\approx1.2$. Thus, elliptical galaxies 
were already well established in X-ray luminous clusters when
the universe was a third of its present age.~
\end{abstract}

\section{Introduction}

Present-day cluster ellipticals are a remarkably well-behaved class of
objects, with structural and chemical properties obeying simple
power-law scaling relations.
But this could not always have been the case in
a hierarchical universe.
While most galaxy formation models can be tuned
to reproduce these relations at $z{\,=\,}0$, 
a more stringent test lies in reproducing their evolution with redshift.
To this end, it is important to study rich clusters out to the highest
redshifts, when fractional age differences among the galaxies were
proportionately greater.
In recent years, deep wide-field optical surveys 
and deep serendipitous X-ray surveys have uncovered significant
numbers of rich galaxy clusters to redshift unity and beyond
(see reviews by Postman 2002; Rosati 2003).
These most distant, and most massive, of known gravitationally bound
structures can then be studied in detail through 
targeted, high-resolution, follow-up optical and near-infrared observations.

We have undertaken a survey of rich galaxy clusters in the
redshift range $0.8<z<1.3$ using the Advanced Camera for Surveys
(ACS; Ford \etal\ 2002) on the {\it Hubble Space Telescope} (\hst).
The aim of this survey is to establish new constraints on the
cluster formation epoch and the evolution of early-type galaxies.
The first cluster observed,
RDCS\,1252.9--2927 (hereafter \clname)
at $z{\,=\,}1.237$ (Rosati 2003; Rosati \etal\ 2003), was discovered 
as part of the ROSAT Deep Cluster Survey (Rosati \etal\ 1998)
and is among the highest-redshift galaxy clusters with
spectroscopic confirmation.  This Letter presents the first results
from our ACS cluster survey, focusing on the color-magnitude (CM)
relation of the early-type galaxies in \clname.
% Unless otherwise noted,
We adopt the best-fit WMAP cosmology:
$(h,\Omega_m,\Omega_L) = (0.71,0.27,0.73)$
(Bennett \etal\ 2003), giving a scale of
8.4 kpc per arcsec at $z{\,=\,}1.237$.

\begin{figure*}
\epsscale{2}
\plotone{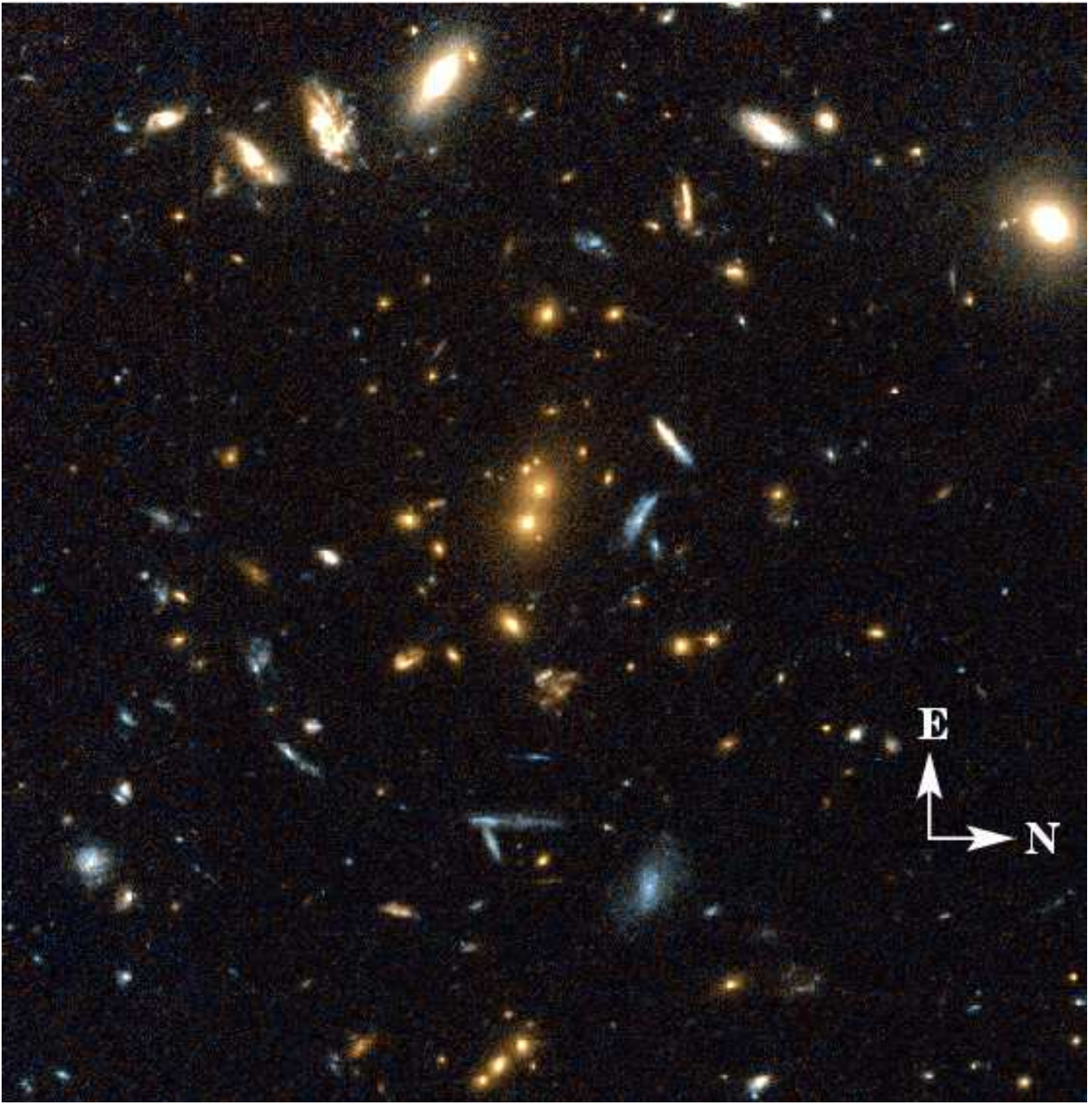}
\bigskip
\caption{Color composite of the core region of \clname, constructed from
our ACS/WFC F775W and F850LP images, shown in the observed orientation.
The displayed field size is roughly 1\arcmin\ across, or less than 4\% of 
the full mosaic.\bigskip\bigskip\bigskip
\label{fig:cluscenter}}
\end{figure*}

\section{Observations and Image Reductions}

\clname\ was observed in the F775W and F850LP bandpasses 
(hereafter \iacs\ and \zacs, respectively)
with the ACS Wide Field Camera as part of the guaranteed 
time observation program (proposal 9290) during 2002 May and 2002 June.  
The observations were done in a $2{\times}2$ mosaic pattern, with 3 and 5 orbits
of integration in \iacs\ and \zacs, respectively, at each 
of the four pointings.  There was nearly 1\arcmin\ of overlap between 
pointings; thus, the core of cluster was imaged for a total of 12 orbits in \iacs\
and 20 orbits in~\zacs. 

The data were processed with the ``Apsis'' pipeline described by
Blakeslee \etal\ (2003), with some recent updates.  In particular, we used
a version of the drizzle software (Fruchter \& Hook 2002) supplied by
R.\,Hook that implements the ``Lanczos3'' interpolation kernel (a damped 
sinc function).  This kernel produces a sharper point spread function
(PSF) and greatly reduces the noise correlation of adjacent pixels
and the resulting ``moire'' patterns.  Apsis also now
removes discontinuities in the residual bias level at the amplifier
boundaries, producing a more uniform background.
An earlier processing of these images has been used by Bouwens \etal\ (2003)
for a study of the faint \iacs\ dropout population at $z{\,\sim\,}6$.
%
% m1i = 25.6405 
% m1z = 24.8432 
% A_i = 0.1508
% A_z = 0.1101
We calibrate our photometry to the AB system using photometric zero points
% for 1$\,e^-\,$sec$^{-1}$ 
of 25.640 (\iacs) and 24.843 (\zacs).
These are uncertain at the $\sim\,$0.02 mag level, 
which has no effect on our conclusions.
We adopt a Galactic reddening for this field of
$E(i{-}z) = 0.041$\,mag based on the Schlegel \etal\ (1998) dust maps.

% \section{Galaxy Selection and Photometry}
\section{Object Selection and Photometry}

Figure\,\ref{fig:cluscenter} shows the central ${\sim}1\arcmin$ region of a 
color composite made from our reduced \iacs\ and \zacs\ images. 
A red galaxy population is clearly visible. The central pair of galaxies
% which are each of magnitude $\zacs\approx21$, are separated by 1\farcs81.
are separated by 1\farcs8 (15\,kpc) and are each of magnitude $\zacs\approx21$.
We used SExtractor (Bertin \& Arnouts 1996) 
in ``dual-image mode'' with low threshold and deblending settings to 
find objects in the reduced images and perform the initial photometry.  
SExtractor ``{\sc mag\_auto}'' values were used for the total magnitudes.  
The \iz\ color effectively separates out evolved 
galaxies at $z\gta1$, and the cluster is obvious as a central
concentration of galaxies with $0.80<\iz<1.05$.
%%%%%%%%% add back in removed fig for astroph
Figure\,1b [removed from the ApJL version in order to meet the
page limit] shows histograms of isophotal color within 3 different
radii of the cluster center (defined midway between the two central
galaxies) for galaxies with total $ \zacs = 20$--25 mag.
%%%%%%%%%%

We selected an initial sample of 312 nonstellar 
objects with $\zacs<24.8$, in the broad isophotal color 
range $0.5<\iz<1.2$, and inside a radius of 1\farcm92.
Our goal is to study the early-type
galaxy population in \clname, for which we have limited
spectroscopic data, and these cuts are designed to select 
the vast majority of our target sample while reasonably
limiting foreground/background contamination.  The color 
selection is roughly 7~times broader than the full-width
of the red sequence we find below.
% and thus fully encompasses the early-type cluster members.
The radial cutoff corresponds to about 1.0 Mpc for both
our adopted WMAP cosmology and
an Einstein-deSitter cosmology with $h{\,=\,}0.5$.
%and an $h{\,=\,}0.5$, $\Omega_m{\,=\,}1$ cosmology.

\par\bigskip\smallskip
\vbox{\baselineskip 9pt
\psfig{file=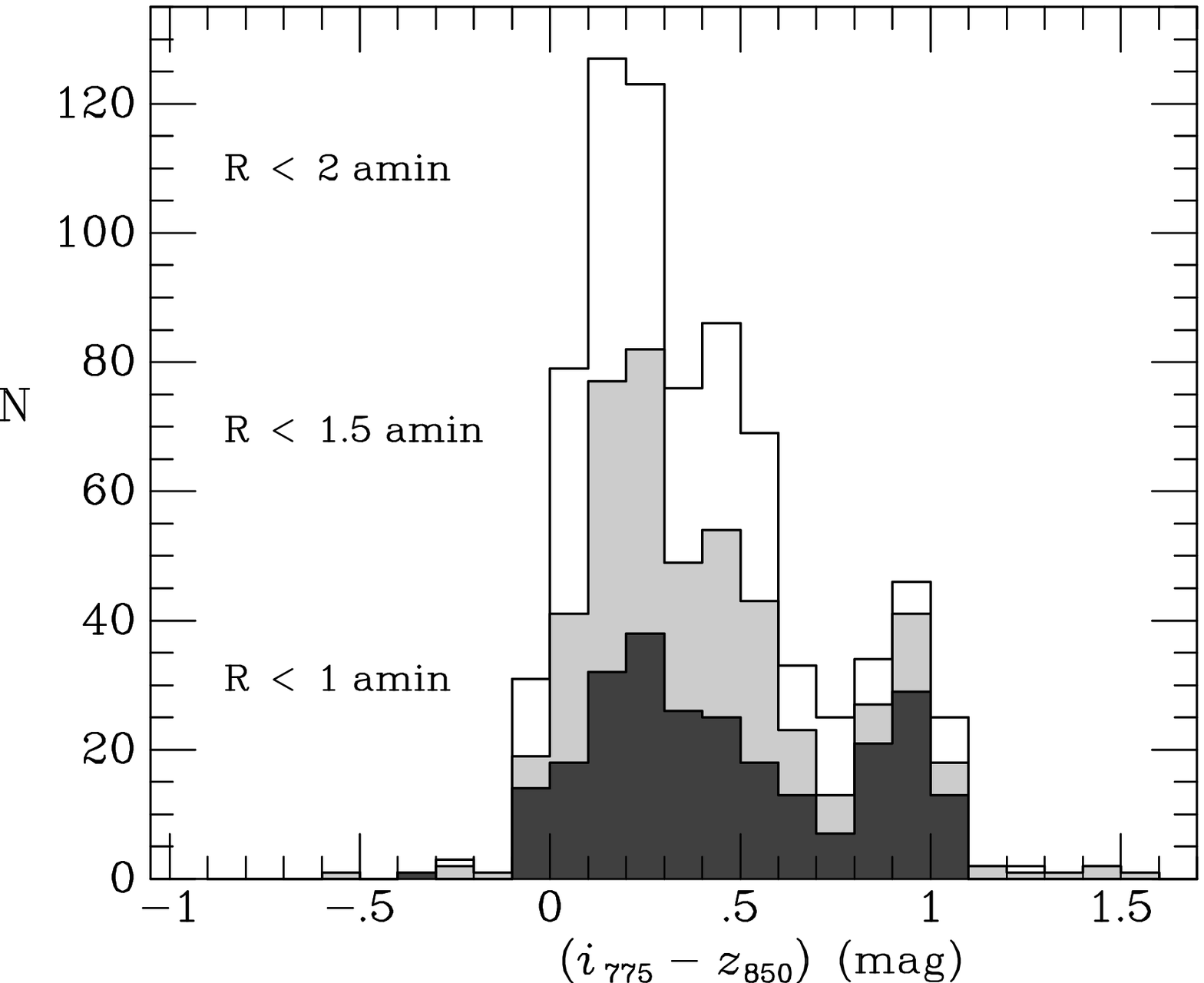,width=8.4cm}
{\par\smallskip\footnotesize
{``{\sc Fig.} 1b.''--- Histograms of \iz\ color for objects  having $20<\zacs<25$
and within the specified radii of the center of \clname.
The cluster galaxies are near $\iz=0.95$, and make up
an increasing fraction of the objects within progressively
smaller radii. [This figure was removed from the published
version because of space limitations.]\vspace{0.5cm}}}}
\par\bigskip\bigskip

Our final colors are measured within galaxy effective 
radii $R_e$ to avoid biasing the CM slope due to color gradients.
We follow the basic approach outlined by van Dokkum \etal\
(1998, 2000).  We derive the $R_e$ values using the program ``galfit''
(Peng \etal\ 2002) by fitting each galaxy to a Sersic model 
(convolved with the PSF),
but constraining the $n$ parameter such that $1\leq n\leq4$.  
Bright neighboring
galaxies were fitted simultaneously.  We note that subtraction 
of the model for the two central galaxies reveals evidence for
interaction in the form of an \textsf{S}-shaped residual.

Next, we  deconvolve
\iacs\ and \zacs\ postage stamp images of each galaxy using the
CLEAN algorithm (H{\"o}gborn 1974) in 
order to remove the differential blurring effects of the PSF,
which is $\sim\,$10\% broader in the \zacs\ band.  To reduce
noise, the CLEAN maps are smoothed with a Gaussian of 
$\hbox{FWHM}{\,=\,}1.5$ pix before adding the residual images to
the maps to ensure flux conservation.  We then measure the
\iacs\ and \zacs\ magnitudes of each galaxy within a circular
aperture of radius $R_e$, typically about 6 pix, or 0\farcs3.
We did not allow the radius to drop below 3~pix.
Photometric errors were determined empirically from the pointing
overlap regions.  We reprocessed the 4 pointings separately and
measured the color differences within $R_e$ for 202 pairs of
measurements for 74 different early-type galaxies 
(classifications described below) in the overlap regions.
We then median-filtered to obtain
the errors as a smooth function of magnitude.
The error thus determined for single-pointing \iz\ 
measurements at $\zacs = 23$ was 0.025\,mag,
rising to $\sim\,$0.05\,mag at $\zacs = 24.5$.  

% \section{Morphology and Membership}

Each galaxy in our initial sample was examined and
morphologically classified, following a procedure similar to
that of Fabricant \etal\ (2000).
This was done independently of the profile fitting, but the
types show a good correlation with Sersic $n$ index.
Here, we simply distinguish between E, S0, and later types, 
where the intermediate S0 class indicates the apparent
presence of a disk without spiral or other structure.
Full details on the classifications for a much larger sample,
including the spatial distribution of the various types, will be 
presented by Postman \etal\ (2003, in preparation).   
About 180 galaxies in this field have measured redshifts,
obtained with VLT/FORS (Rosati \etal\ 2003, in preparation), 
with 31 being cluster members.  
All galaxies in our initial sample classified as early-type,
and not known to be interlopers from their spectra, are
included in our CM analysis in the following section.  Of the 31
known members, 22 are classified as early-type, 
and we include all of these in our analysis 
% all of which we include in our analysis 
even though one (an S0) happened to lie beyond our 1~Mpc cutoff.~

\section{The Color-Magnitude Relation at $z{\,=\,}1.24$}

% talk about:
%  fitting, slope+zeropoint, late-types, scatter, inferred age

%
We fitted the early-type galaxy CM relations
using simple linear least squares; other methods gave very
similar results. We estimate the scatter from both the standard
rms and the biweight scale estimator (Beers \etal\ 1990).
No rejection was done in fitting subsamples
composed of known members, the faintest of which has
$\zacs=23.48$, or  ${\sim\,}0.5\,L^*_B$.
 We also performed
fits to samples with unconfirmed members, allowing us to go 
1\,mag further down the galaxy luminosity function.
% but here we iterate to reject the 3-$\sigma$ outliers.
However, here we iterate to reject the 3-$\sigma$ outliers,
as these are likely to be interlopers: none of the 22 confirmed 
early-type members is more than 2.3-$\sigma$ discordant.
After the iterative rejection process, 
we find concordant scatters for those samples, and
the rms and biweight estimator
% yield virtually identical results.
 are the same to within $\pm$0.001\,mag.
%%
%   zpt    +/-      slope    +/-  Ndat   rmssig   bwtsig bestsig
%  1.5317 0.1492  -0.02493 0.00641   52  0.03787  0.03857 0.03822  early
%  1.3886 0.1717  -0.01874 0.00743   31  0.03658  0.03643 0.03643  ellip
%  1.3673 0.2931  -0.01753   0.013   22  0.0384   0.03819 0.03819  earlymemb
%
% recentering, z-23:
%    zpt      +/-      slope    +/-  Ndat   rmssig   bwtsig bestsig  lim
% 0.95829 0.005482  -0.02493 0.00641   52  0.03787  0.03857 0.03822  24.5
% 0.95806 0.005869  -0.02537 0.00825   41  0.03756  0.03768 0.03762  24.0

Figure~\ref{fig:finecmr} presents the CM relation for the 
\clname\ galaxies.  The fit to the full sample of early-type
galaxies with $\zacs<24.5$ gives
\begin{equation}
%\iz = (1.532\pm0.149) - (0.025\pm0.006)\,\zacs\,, \label{eq:allearly}
%\iz = (0.958\pm0.006) - (0.025\pm0.006)(\zacs-23)\, . \label{eq:allearly}
(i{-}z) \,= \,(0.958\pm0.006) - (0.025\pm0.006)(\zacs-23)\, . \label{eq:allearly}
%\iz \,= \; 0.958 \,-\, 0.025 (\zacs-23)\, . \label{eq:allearly}
\end{equation}
Other results are listed in Table~\ref{tab:cmr}.  
The mean locations of the CM
relations for the elliptical and S0 subsamples agree to well within
the errors, while the slopes are consistent at the 1.5-$\sigma$ level.  
Eight known late-type members from Rosati \etal\ 
(2003, in preparation) for which we have photometry
are bluer than the early-type galaxies by 0.25\,mag, with 
a scatter of 0.14 mag about this offset, indicating young
stellar populations.
We find an intrinsic scatter  $\sigma_{\rm int} = 0.023\pm0.007$
mag for the 15 confirmed elliptical members.
For a limit of $\zacs < 24.5$, we derive
$\sigma_{\rm int} = 0.026$ mag for the clipped sample 
of 31 ellipticals and $\sigma_{\rm int} = 0.029$
for the 52 E+S0 galaxies.  
At this limit, the observational errors become
dominant and classification is difficult, which could
bias our $\sigma_{\rm int}$ estimates.  For $\zacs < 24.0$, 
still 3\,mag down the luminosity function to about 0.3\,$L^*$,
we find $\sigma_{\rm int} = 0.024$ for 25 ellipticals
(with no outliers).

\par\bigskip\smallskip
%\vbox{\vspace{8pt}
\psfig{file=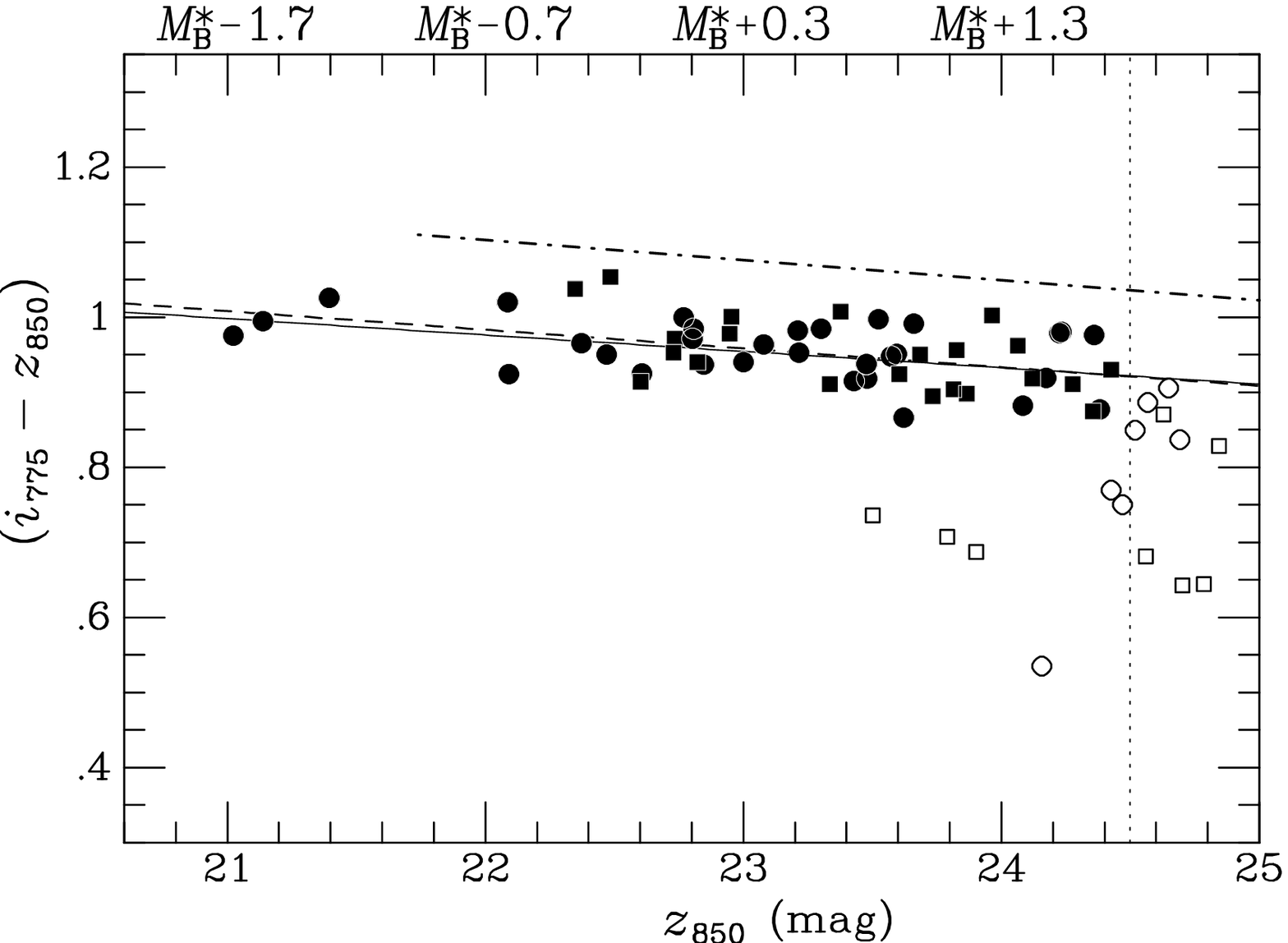,width=8.7cm}
\figcaption{Color-magnitude diagram for early-type
galaxies inside $1\farcm9$ with $\iz>0.5$, excluding 
spectroscopically known interlopers.
% as well as galaxies classified as late-type
% and not known to be in the cluster.
Circles and squares represent ellipticals and S0s,
respectively; solid symbols are used in the CM relation
fits, while open symbols (all of which lack spectroscopic
information) are rejected as outliers or as
below the faint cutoff (indicated by the dotted line). 
% Finally, open stars are confirmed late-type members.
Two representative fits are shown: the  fit to the 15
elliptical members (solid line) and to the 52 early-type
red-sequence galaxies. 
% The labels at top give the approximate luminosity conversion,
The approximate luminosity conversion for \clname\ is shown at top,
assuming the WMAP cosmology and $-1.4$\,mag of luminosity evolution
as described in the text, such that $M_B^* = -21.7$ (AB).
The relation for the Coma cluster, transformed
to these bandpasses at $z=1.24$ (no evolution correction),
is indicated by the dot-dashed line.
\label{fig:finecmr}}
\par\bigskip

We estimate $M_B^* \approx -21.7$ AB for \clname, based on
local surveys (Norberg \etal\ 2002),
WMAP cosmology, and $-$1.4 mag luminosity evolution
(Postman \etal\ 2001; van Dokkum \& Stanford 2003).
The correction from \zacs\ to rest-frame $B$ is $+$0.3 mag
for an early-type spectrum.  Thus the brightest cluster
member at $\zacs=21.0$ corresponds to ${\sim\,}4.8\,L^*_B$.
The cluster red-sequence is about 0.1\,mag 
bluer than predicted for non-evolving elliptical templates
(Coleman \etal\ 1980; Schneider \etal\ 1983).
For comparison, a Bruzual \& Charlot (2003; hereafter BC03)
solar metallicity model reddens by 0.11 mag and fades by 1.5\,mag
in aging from $\sim\,$3 to $\sim\,$11.5 Gyr.
The present-day relation for the Coma cluster (transformed according
to the relations given below) is also shown in Figure~\ref{fig:finecmr}.

The tight CM relation allows us to constrain the scatter
%early-type galaxy ages, for a given model of the color evolution.
in early-type galaxy ages, subject to model uncertainties.
The evolution in \iz\ at this redshift is
complicated at young ages because \zacs\ straddles the 
4000\,\AA\ break and has its blue end near the Balmer jump
at 3700\,\AA.  The Balmer jump
reaches a maximum at 0.5--1 Gyr, when the relative
contribution from A stars is greatest, resulting in a
red color at these ages.  It is interesting that
% a few of the S0s and one later type galaxy lie {\em above}
several of the S0s lie {\em above}
the CM relation, possibly indicating ages ${<\,1}$ Gyr.
As the A~star contribution
lessens, the color quickly evolves towards the blue,
reaching a local minimum near $\sim\,$1.5 Gyr, before
commencing a roughly monotonic reddening with age.

We have simulated the \iz\ colors of galaxies formed under two simple
star formation models, similar to those used by van Dokkum \etal\ (1998).
In Model~1, the galaxies form in single bursts randomly distributed
over the interval $(t_0,t_{\rm end})$, where $t_0$ is set to the recombination
epoch and $t_{\rm end} < t_z\equiv 5.1$ Gyr, the age of the universe at $z=1.24$.
In Model~2, galaxies form stars at constant rates between
randomly selected times $(t_1,t_2)$, where $t_0<t_1<t_2<t_{\rm end}$.
We vary $t_{\rm end}$ and at each step calculate colors and
luminosities for 10,000 ``galaxies'' by interpolation and integration
of the BC03 solar metallicity models.
%, rather than power-law approximations as in van Dokkum et~al.
Assuming $\sigma_{\rm int} =0.024$ mag for the ellipticals, 
Model~1 imples a minimum age $t_z{-}t_{\rm end} = 1.6$ Gyr,
i.e., all galaxies finish forming at redshifts $z>z_{\rm end}=1.9$;
the mean luminosity-weighted age is $\taul = 3.3$ Gyr 
(corresponding to $z_L=3.6$) with a scatter of 30\%.
Model~2 gives $t_z{-}t_{\rm end} = 0.53$ Gyr, $z_{\rm end}=1.4$, 
and $\taul = 2.6$ Gyr (corresponding to $z_L=2.7$) with 38\% scatter.
Thus, although some galaxies in Model~2 have formed stars recently,
the mean ages are still high.
Both models give a mean color $\langle \iacs{-}\zacs\rangle = 0.94$,
similar to that observed.

For the S0s, we find $z_{\rm end}=1.5$ for Model 1 and $z_{\rm end}=1.3$, indicating 
recent star formation, and age scatters of 44--47\%.
Finally, we note that the 1996 version of the BC models would
have predicted higher formation epochs, e.g., $z_{\rm end}>2.5$ and $z_{\rm end}>2.0$ for 
the ellipticals in Models 1 and 2, respectively, and an age scatter of only 20\%,
although the predicted color is then redder by 0.1 mag.
Overall, we conclude that the ellipticals are an evolved population,
with a mean age $\taul\gta2.6$, a minimum age ${\sim\,}1\pm0.5$ Gyr,
and an age scatter of ($34\pm15$)\%,
where the error reflects
uncertainty in $\sigma_{\rm int}$ and scatter in the models.

\section{Discussion}

% Slope, scatter, zeropoint compared to local universe.

To enable comparison with previous work, we convert observed
% In order to compare to previous work, we convert observed
\izacs\ quantities to rest-frame \ubrest\ at $z = 1.24$.
The models (BC03; Kodama \etal\ 1998) and empirical templates
indicate $\Delta \ubrest = (1.8\pm0.4){\,\times\,}\Delta$\izacs, 
where the error bar reflects the scatter in the models
at the relevant ages and adds about 20\% uncertainty
to our transformed slope and scatter. Figure~\ref{fig:scatterz}
uses this conversion, and other transformations from 
van Dokkum \etal\ (2000), to compare our results to some previous
studies of the CM relation in intermediate-redshift
clusters with \hst, as well as the results of
van Dokkum \etal\ 2001 on RX\,J0848+4453, the only 
cluster of comparable redshift to have its CM relation
measured.  

% We find no compelling evidence for evolution in either
% the slope or scatter of the \ubrest\ CM relation for cluster ellipticals.

Linear fits to the data shown in Figure~\ref{fig:scatterz}
yield slopes of $-0.014\pm0.010$ and $0.003\pm0.008$ for the
evolution in the absolute slope and the scatter, respectively.
% This is consistent with our findings in the previous section
% that ellipticals in \clname\ are already an evolved population.
Thus, the scatter is constant, and there is at best marginal
%evidence for slope evolution, indicating that the
evidence for slope evolution, which indicates that the
slope is due to a variation in metallicity, not age.
Previous studies of cluster samples out to $z\sim1$
have come to similar conclusions
(e.g., Stanford \etal\ 1998; Kodama \etal\ 1998; van Dokkum \etal\ 2000).
Van Dokkum \etal\ (2001) concluded that the slope at $z=1.27$ was
shallower than in the Coma cluster.
However, as shown in the figure, our slope measurement is consistent
within the errors with both Coma and RX\,J0848+4453.  Further
studies of a diversity of clusters at similar redshifts are needed to
explore this issue.

The lack of evolution in the CM relation scatter can be
explained by progenitor bias (e.g., van Dokkum \& Franx 2001): galaxies
selected as early-type at any epoch will have old stellar
populations, while the later-type progenitors of the youngest
ellipticals today will not be selected.  
The result is an underestimate in the color scatter for the progenitors 
of modern ellipticals, and thus overestimated ages.
An upper limit on the scatter for elliptical progenitors 
may be estimated from a fit to all confirmed \clname\ members; 
% Table~\ref{tab:cmr} shows this is
the result is 3--4 times larger than for the early-type galaxies.
A detailed study of the morphological fractions in \clname\
(Postman \etal\ 2003, in preparation) 
should help illuminate the magnitude of this bias.
We also note that the two central ellipticals themselves,
based on their proximity and irregular isophotes, 
appear likely
to undergo dissipationless merger to form a single ${\sim\,}9\,L^*_B$ galaxy,
similar to local cD galaxies.~~

We conclude that massive, evolved early-type galaxies were already 
present in rich clusters at $z=1.24$.
Our simple models imply mean luminosity-weighted ages of 2.6--3.3 Gyr, 
corresponding to formation at $z=2.7$--3.6.
However, the \iz\ color is not ideal for the redshift of \clname, being
better suited to measuring the 4000\AA\ break at $z\approx1.1$.  
Combining our ACS data with
deep, high-resolution near-IR imaging of
this field (Lidman \etal\ 2003) will enable a more robust assessment of early-type
galaxy ages.  In addition, further studies of the CM
relations and morphologies of galaxies in other $z\gta1$ clusters 
are needed to improve the constraints on the formation epoch of cluster
galaxies and on the evolution of their stellar
populations and structural properties.

\acknowledgments 

ACS was developed under NASA contract NAS 5-32864, and this research 
has been supported by NASA grant NAG5-7697.
The {STScI} is operated by AURA Inc., under NASA contract NAS5-26555.
We thank our fellow ACS Team members for their help,
Taddy Kodama for helpful discussions and models,
and Stephane Charlot for the BC03 models.

\par\vbox{\vspace{0.7cm}
\psfig{file=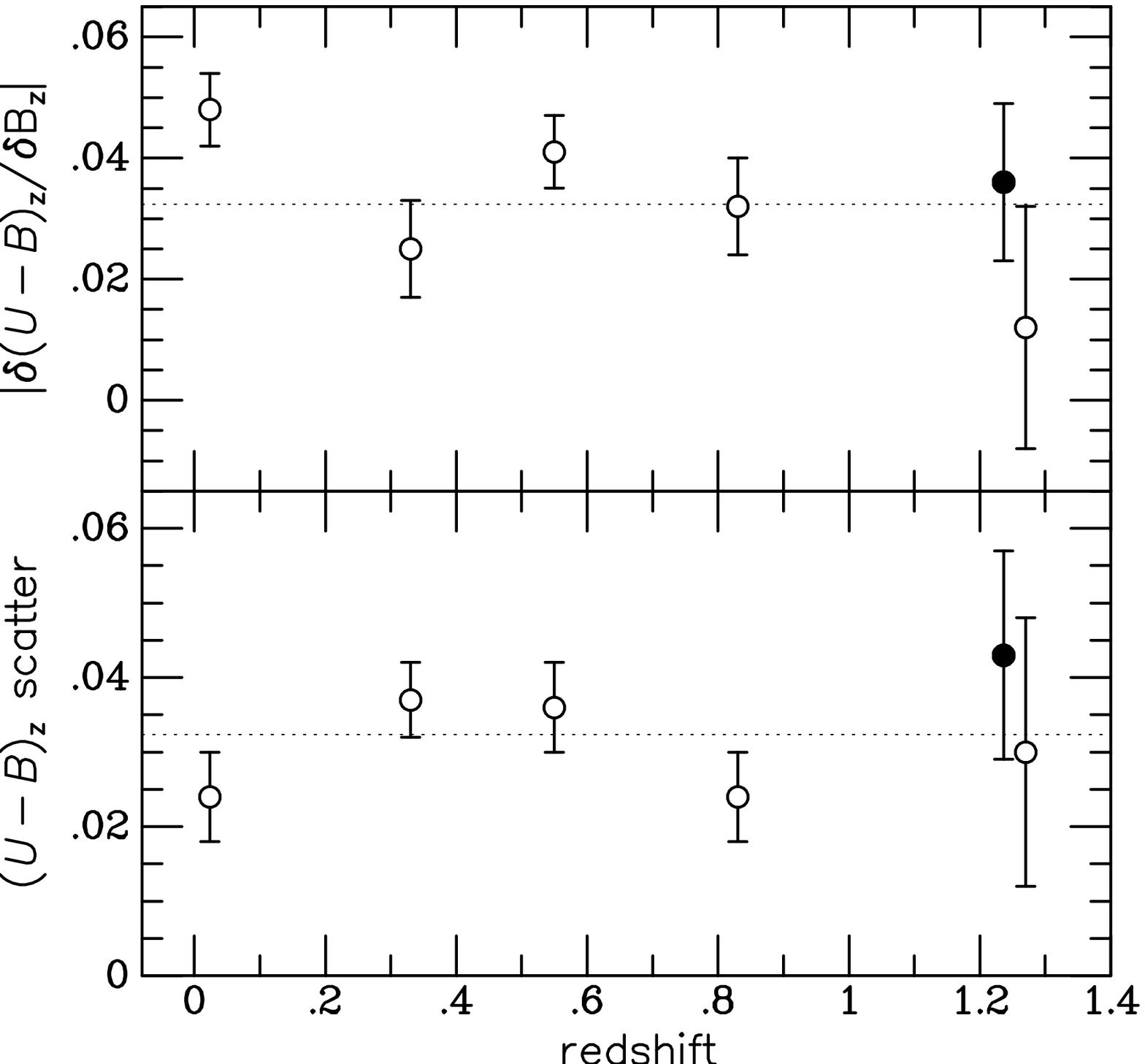,width=8.3cm}\figcaption{
Slope (top) and scatter (bottom) of the rest-frame (\ub)
CM relation as a function of redshift. The filled circle is 
from the present work; open symbols show results from,
in order of increasing redshift, Bower \etal\ (1991); 
van Dokkum \etal\ (1998); Ellis \etal\ (1998); 
van Dokkum \etal\ (2000); and van Dokkum \etal\ (2001). 
The dotted lines indicate the average values.
\label{fig:scatterz}}
}\par

\begin{deluxetable}{lcccccc}
\vspace{-20pt}
 \tabletypesize{\small}
% \tabletypesize{\scriptsize}
% \tabletypesize{\footnotesize}
\tablewidth{0pt}   % 0pt sets table to its natural width
\tablecaption{RDCS\,1252-2927 Color-Magnitude Relations}
\tablehead{
\colhead{Sample} &
\colhead{$z_{850}^{\rm (lim)}$} &
\colhead{$N_{\rm r}$ ~ $N_{\rm c}$} &
\colhead{Slope} &
\colhead{$\sigma_{\rm bwt}\tablenotemark{a}$} &
\colhead{$\sigma_{\rm bwt}\tablenotemark{b}$} &
\colhead{$\sigma_{\rm int}$}
}
\startdata
E\tablenotemark{c}\dotfill    &\dots &  15 ~ 15 & $-0.022\pm0.011$ & 0.029 & $0.029\pm0.005$ & $0.023 \pm 0.007$\\
E+S0\tablenotemark{c}\dotfill &\dots &  22 ~ 22 & $-0.018\pm0.013$ & 0.038 & $0.038\pm0.006$ & $0.033 \pm 0.007$\\
E\tablenotemark{d}\dotfill    & 24.0 &  25 ~ 25 & $-0.020\pm0.009$ & 0.033 & $0.033\pm0.005$ & $0.024 \pm 0.008$\\
E+S0\tablenotemark{d}\dotfill & 24.0 &  44 ~ 41 & $-0.025\pm0.008$ & 0.045 & $0.038\pm0.004$ & $0.029 \pm 0.007$\\
E\tablenotemark{d}\dotfill    & 24.5 &  34 ~ 31 & $-0.019\pm0.007$ & 0.053 & $0.036\pm0.005$ & $0.026 \pm 0.008$\\
E+S0\tablenotemark{d}\dotfill & 24.5 &  58 ~ 52 & $-0.025\pm0.006$ & 0.054 & $0.039\pm0.004$ & $0.029 \pm 0.007$\\
S0\tablenotemark{d}\dotfill   & 24.5 &  24 ~ 21 & $-0.042\pm0.013$ & 0.058 & $0.039\pm0.006$ & $0.032 \pm 0.008$\\
% it's correct, to this precision,  E+S0 sample gives identical results
% for mcut =24.0,24.5
\enddata
\vspace{-6pt}
\tablenotetext{a}{Biweight scatter based on raw number of galaxies $N_{\rm r}$.}
\tablenotetext{b}{Biweight scatter based on $N_{\rm c}$ galaxies after 3-$\sigma$ clipping.}
\tablenotetext{c}{Spectroscopically confirmed members of specified type only.\vspace{-3pt}}
\tablenotetext{d}{All red-sequence objects (known interlopers omitted) of specified type,
 $\zacs < z_{850}^{\rm (lim)}$ and within the area of analysis.}
\label{tab:cmr}
\end{deluxetable}

\end{document}